\documentclass[10pt]{article}
\usepackage{extsizes}
\usepackage[utf8]{inputenc}
\usepackage[english]{babel}

\usepackage[T1]{fontenc}

\usepackage[font=footnotesize]{caption}

\usepackage{amsmath}
\usepackage{dirtytalk}
\usepackage{scalerel}
\usepackage{slashed}
\usepackage{setspace}
\usepackage{amsfonts}
\usepackage{bigints}
\usepackage{mathtools}
\usepackage{float}
\usepackage{amssymb}

\usepackage{graphicx}
\usepackage{upgreek}
\usepackage{parskip}
\usepackage{bigints}
\usepackage{multicol}
\usepackage{multirow}

\renewcommand\arraystretch{0.7}

\newcommand\numberthis{\addtocounter{equation}{1}\tag{\theequation}}

\usepackage[hidelinks]{hyperref}
\usepackage{hyperref}

\usepackage{sectsty}
\sectionfont{\fontsize{9}{15}\selectfont\centering}

\usepackage[sort&compress,square,numbers]{natbib}
\usepackage[margin=0.6in]{geometry}
\usepackage{parskip}
\usepackage{tabularx}
\usepackage{longtable}
\usepackage{mleftright}

\usepackage[dvipsnames]{xcolor}
\usepackage{xcolor}
\newcommand\tcolor[1]{\textcolor{Blue}{#1}}

\newcommand{\ga}{\scaleobj{1.2}{\gamma}}
\newcommand{\Tr}{\textnormal{Tr}}
\newcommand{\su}[1]{u\scaleobj{0.8}{(#1)}}
\newcommand{\sub}[1]{\overline{u}\scaleobj{0.8}{(#1)}}

\newcommand{\sv}[1]{v\scaleobj{0.8}{(#1)}}
\newcommand{\svb}[1]{\overline{v}\scaleobj{0.8}{(#1)}}

\newcommand{\spa}{\hphantom{.}}
\newcommand{\esp}{\enskip}

\newcommand{\comment}[1]{}
\newcommand{\eu}{\scaleobj{1.3}{e}}

\makeatletter
\newcommand\sixteen{\@setfontsize\sixteen{17pt}{6}}
\renewcommand{\maketitle}
{\bgroup\setlength{\parindent}{0pt}
\begin{center}
\sixteen\bfseries \@title
\medskip
\end{center}
\textit{\@author}
\egroup}
\makeatother

\title{
    {\Large
        \textbf{Comment on \say{Spin correlations in elastic e$^{+}$e$^{-}$ scattering in QED}}
    }
}
\date{}\author{}

\begin{document}
\maketitle

\begin{center}
{Kort Beck} \\
{		
\textit{Department of Physics, University of Illinois at Urbana-Champaign, Urbana, IL 61801, USA\\
Facultad de Ciencias Físico-Matemáticas, Universidad Autónoma de Coahuila, 25000, Saltillo, Coahuila, México}\\
\tcolor{\href{mailto:kgr6@illinois.edu}{kgr6@illinois.edu}}
}
\end{center}

\begin{center}
{Gabriel Jacobo} \\
{	
\textit{Department of Physics, University of Oregon, Eugene, OR 97403, USA\\
Facultad de Ciencias, Universidad de Colima, 28045, Colima, Colima, México}\\
\tcolor{\href{mailto:gjacobo2@uoregon.edu}{gjacobo2@uoregon.edu}}
}
\end{center}



\par\noindent\rule{\textwidth}{0.1pt}
\hphantom{ -- PHANTOM SPACE -- }\\
{\small
In the previous work, ``Spin correlations in elastic $e^+ e^-$ scattering in QED'' (Yongram, 2018), spin correlations for entangled electrons and positrons as emergent particles of electron-positron scattering (also known as Bhabha scattering) were calculated at tree level in QED. When trying to reproduce the author's work, we have found different results. In this work, we show the calculation for fully (initial and final polarized states) and partially (just final polarized states) polarized probability amplitudes for electron-positron scattering at all energies. While Yongram claims that violation of Clauser-Horne inequality (CHI) occurs at all energies for both mentioned cases, for fully polarized scattering we found violation of the CHI for speeds $\beta \gtrsim 0.696$, including the high energy limit, supporting the agreement between QED and foundations of quantum mechanics. However, for initially unpolarized particles we found no violation of the CHI.} 
\par\noindent\rule{\textwidth}{0.1pt}\\


\setlength{\jot}{13pt}
\begin{multicols}{2}

\section{INTRODUCTION}

One of the main differences between a classical and a quantum theory is that, in a quantum theory classical forbidden phenomena might happen such as tunneling and quantum entanglement, the later first studied in 1935 by Einstein, Podolsky and Rosen \tcolor{\cite{EPR}}. Suppose we have two entangled quantum systems $A$ and $B$. If we measure a property of one system (for instance, spin), due to entanglement we also determine automatically the same property of the other system no matter the distance between both systems. This phenomena suggested that quantum mechanics might be an incomplete theory, lacking ``hidden variables''. In 1964, John Stewart Bell introduced the now so-called Bell inequality \tcolor{\cite{Bell}}. This inequality is a bound over correlations of experimental results, violation of the inequality supports the fact that quantum mechanics is a complete theory without hidden variables (at least the ones which fulfils the assumptions of the Bell’s theorem). The original Bell inequality is not optimal to test experimentally \tcolor{\cite{bell_ineq_at_LEP}}, because of this, other Bell-type inequalities have been searched for. Since the Bell-type inequalities were introduced, several experiments have confirmed their violation \tcolor{\cite{BIviol_1,BIviol_2,BIviol_3,BIviol_4,BIviol_5}}, supporting quantum theory as a true description of physical reality.

Quantum entanglement has been broadly studied as a tool to develop new technologies to improve quantum computing. Also, it has been studied as a fundamental nature phenomenon, for example in the context of foundations of quantum mechanics \tcolor{\cite{foundations_qm}} and Quantum Field Theories \tcolor{\cite{qft_ent_1,qft_ent_2,qft_ent_3}}, and even while studying neutrino oscillations \tcolor{\cite{NeutrinoEnt,Neutrino1,Neutrino2,Neutrino3,Neutrino4,Neutrino5}}. Here we are interested in studying entanglement in scattering processes from QED. Recent work on this and related tests of Bell-type inequalities at colliders can be found in \tcolor{\cite{wittness,EntangEntropyDecay,max_ent_DIS,BI_collider_1,moller_scatt,scatt_qed_ent,two_qbit_entang,BI_collider_2,BI_collider_3}}.

In order to determine if a theory belongs to a local hidden variable class, we can test the CHI derived by Clauser and Horne in \tcolor{\cite{CH1974,CH1978}}. The CHI is a Bell-type inequality easier to test experimentally compared to the original Bell inequality. In their work \tcolor{\cite{CH1974}}, Clauser and Horne also proposed an experimental setup to correctly test their inequality. Let

\begin{align*}
S &= \frac{p_{12}(a_1,a_2)}{p_{12}(\infty,\infty)}
- \frac{p_{12}(a_1,a_2')}{p_{12}(\infty,\infty)} 
+ \frac{p_{12}(a_1',a_2)}{p_{12}(\infty,\infty)}\\
&+ \frac{p_{12}(a_1',a_2')}{p_{12}(\infty,\infty)} 
- \frac{p_{12}(a_1',\infty)}{p_{12}(\infty,\infty)} 
- \frac{p_{12}(\infty,a_2)}{p_{12}(\infty,\infty)} \spa\spa,
\numberthis
\end{align*}

if $S$ gets a value outside the range [-1,0], then it is said that the inequality is violated. In this case, $p_{12}(a_1,a_2)/p_{12}(\infty,\infty)$ represents the joint probability of measuring the spin of the outgoing particles along the directions $a_1$ and $a_2$, and $p_{12}(\infty,a_2)/p_{12}(\infty,\infty)$ is the probability of measuring the spin of only one particle in the direction $a_2$. Here, $p_{12}(\infty,\infty)$ denotes the normalization factor. 

In previous work \tcolor{\cite{yongram}}, spin correlations for particles oriented as the ones described in the experimental setup proposed by Clause and Horne were calculated. The probability amplitudes derived depend explicitly on the speed $\beta$ and polarization angles $\chi_1$ and $\chi_2$ for emergent particles of electron-positron scattering. Although authors in \tcolor{\cite{yongram}} claim to have found violation of the CHI at all energies, we found this is not true. Also, we made some corrections on the spinor expressions they used. To verify this, it is enough to take the spinors and plug them into the Dirac equation. On appendices A and B we show the correct expressions for all spinors. 

The work is distributed as follows: (1), first we perform the calculations for initially polarized particles and test the CHI at all energies. (2) Then we repeat the same procedures as before but with initially unpolarized particles, so we averaged over initial spins. Here we also test CHI at all energies, in this case we do not find violation of the CHI. (3) Finally, we discuss possible reasons why CHI violation does not occur for initially unpolarized scattering.


\section{INITIALLY POLARIZED PARTICLES}

\begin{figure}[H]
\includegraphics[scale=0.24]{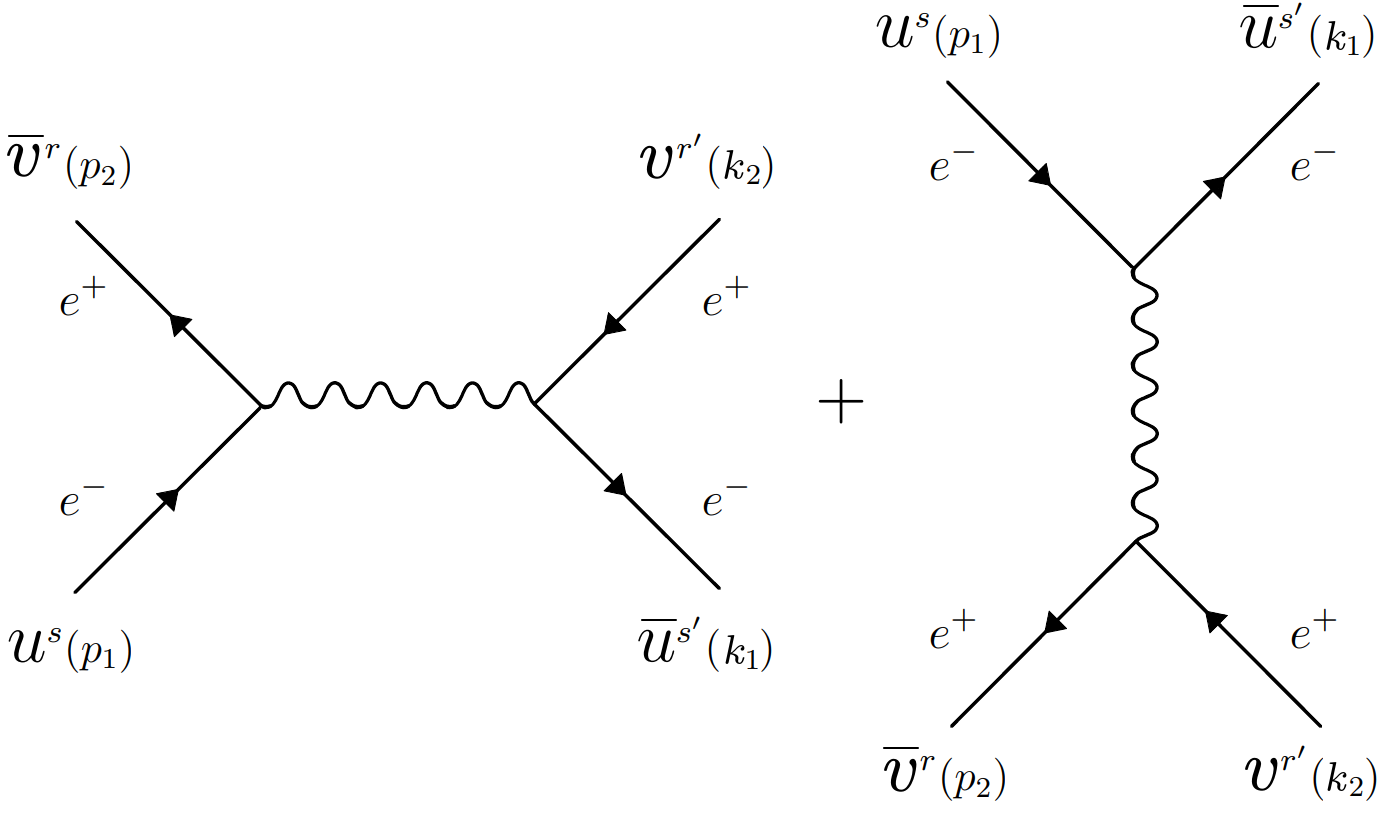}
\centering
\caption{Feynman diagrams for electron-positron scattering.}
\label{feyndiag}
\end{figure}

As already mentioned, the scattering amplitude is calculated a tree level in QED only. The Feynman diagrams are shown in Figure \tcolor{\ref{feyndiag}}.

The probability amplitude for this scattering process shown in Figure \tcolor{\ref{feyndiag}} is known to be

\begin{align*}  \label{general_amplitude}
\mathcal{M} &=
\frac{e^2}{s}  \esp \sub{k_1} \ga^\mu \sv{k_2} \svb{p_2} \ga_\mu \su{p_1} \\
&- \frac{e^2}{t} \esp \sub{k_1} \ga^\nu \su{p_1} \svb{p_2} \ga_\nu \sv{k_2}
\spa\spa. \numberthis
\end{align*}

\begin{figure}[H]
\includegraphics[scale=0.35]{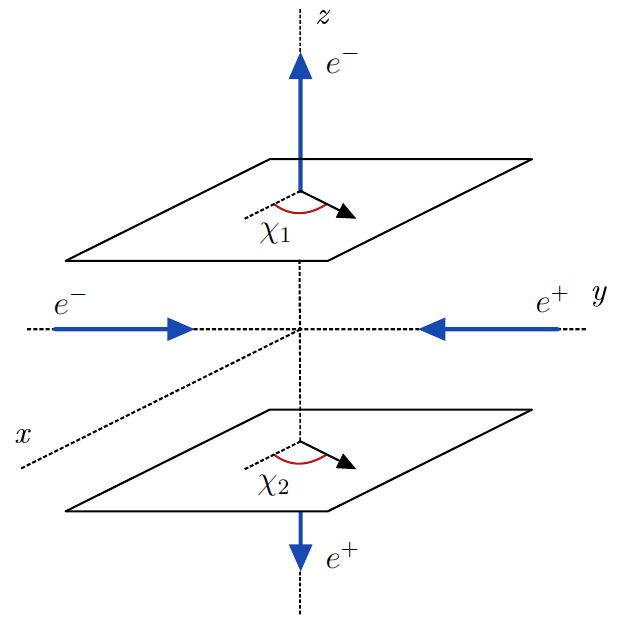}
\centering
\caption{Configuration space scattering diagram. Initial electron and positron move along the $y$ axis, and the emergent electron and positron along the $z$ axis. The angle $\chi_1$ is measured relative to the $x$ axis and denotes the electron spin orientation over the $xy$ plane. The angle $\chi_2$ is also measured relative to the $x$ axis and denotes the positron's opposite spin orientation over the $xy$ plane. Remember that for antiparticles the spinor direction is opposite to physical particle's spin direction.}
\label{scatter_diagram_initially_polarized}
\end{figure}

As done in \tcolor{\cite{yongram}}, we consider the particles' momentum and polarization oriented as in the experimental setup proposed by Clauser and Horne in \tcolor{\cite{CH1974}}. The initial electron and positron are moving along the $y$ axis and colliding in their common CM reference frame. Each particle has spin up in the $z$ direction and momentum $\mathbf{p}_1=\gamma m \beta(0,1,0)=-\mathbf{p}_2$ respectively, where $\gamma=1 / \sqrt{1-\beta^2}$. For the emerging electron and positron, we considered the momenta $\mathbf{k}_1=\gamma m \beta(0,0,1)=-\mathbf{k}_2$ respectively. Using the correct spinor expressions shown in appendix A for the initial and final particles, we find the invariant amplitude to be 

\begin{align*} \label{M_amplitude_section_2}
\mathcal{M} &= \enskip \frac{e^2(\beta\rho - 2)}{4\beta\rho} \enskip
\Bigg[ \spa 
A(\beta) \cos{\left(\frac{\chi_1 +\chi_2}{2}\right)}  \\
&+ B(\beta) \sin{\left(\frac{\chi_1 -\chi_2}{2}\right)} 
- i C(\beta)\sin{\left(\frac{\chi_1+\chi_2}{2}\right)}
\spa \Bigg] 
\spa\spa, \numberthis
\end{align*}

where $\chi_1$ and $\chi_2$ are spin directions over the $xy$-plane specified in Figure \tcolor{\ref{scatter_diagram_initially_polarized}}, and

\begin{align*}
A(\beta) &= (\beta^2 - 1)(\rho^2 + 1)^2 \spa\spa,  \\
B(\beta) &= 8\rho^2 \spa\spa,  \\  
C(\beta) &= (\beta^2 - 1)(\rho^4 - 1) \spa\spa, \\ 
\rho(\beta) &= \frac{\gamma \beta}{\gamma +1}=\frac{\beta}{1+\sqrt{1-\beta^2}}  
\spa\spa. \numberthis
\end{align*}

Let $F(\beta,\chi_1,\chi_2)$ be the modulus squared of $\mathcal{M}$ in equation (\ref{M_amplitude_section_2}), then the conditional joint probability of measuring the polarization of the emergent particles in the directions $\chi_1$ and $\chi_2$ is 

\begin{equation}
P(\beta,\chi_1,\chi_2) = \frac{F(\beta,\chi_1,\chi_2)}{N(\beta)}  \spa\spa.
\end{equation}

The normalization factor $N(\beta)$ is obtained by summing the non-normalized probability over the pair of angles $(\chi_1, \chi_2)$, $(\chi_1+\pi, \chi_2)$, $(\chi_1, \chi_2+\pi)$ and $(\chi_1+\pi, \chi_2+\pi)$, this is

\begin{align*}
N(\beta) &= 
F(\beta,\chi_1,\chi_2) 
+ F(\beta,\chi_1 +\pi ,\chi_2) \\
&+ F(\beta,\chi_1,\chi_2 +\pi )
+ F(\beta,\chi_1 +\pi ,\chi_2 +\pi)  \\
&=  \frac{2e^4}{\beta^4} \spa \big( 2-5\beta^2 +8\beta^4 -\beta^6 \big) 
\spa\spa.
\numberthis
\end{align*}

The probability of measuring only the spin of the electron in the $\chi_1$ direction is

\begin{align}
P(\chi_1,-) &= P(\chi_1,\chi_2)+P(\chi_1,\chi_2+\pi) \spa\spa \notag, \\
&=\frac{1}{2} - \frac{2\beta^2 (\beta^2 - 1 )\sin(\chi_1)}{\beta^6 -8 \beta^4 + 5\beta^2 - 2} 
\spa\spa ,
\end{align}

similarly, the probability of measuring only the spin of the positron in the $\chi_2$ direction is

\begin{align}
P(-,\chi_2) &= P(\chi_1,\chi_2)+P(\chi_1 +\pi,\chi_2) \spa\spa\notag, \\
 &= \frac{1}{2} + \frac{2\beta^2 (\beta^2 - 1)\sin(\chi_2)}{\beta^6 -8 \beta^4 + 5\beta^2 - 2} 
\spa\spa.
\end{align}

If we take the high energy limit $\beta \longrightarrow 1$ the probability becomes

\begin{align}
P(\chi_1,\chi_2) = \frac{1}{4} \Big( 1 - \cos (\chi_1-\chi_2)  \Big)
\spa\spa.
\end{align}

Using these results we can construct the quantity $S$ following Clauser and Horne in \tcolor{\cite{CH1974}} 

\begin{align*}  \label{S_section_2_fully_polarized}
S(\beta) &= P(\beta,\chi_1,\chi_2) - P(\beta,\chi_1,\chi'_2) + P(\beta,\chi'_1,\chi_2) \\
&+ P(\beta,\chi'_1,\chi'_2) - P(\chi'_1 , - ) - P( - , \chi_2 )
\spa\spa.
\numberthis
\end{align*}

\begin{figure}[H]
\includegraphics[scale=0.21]{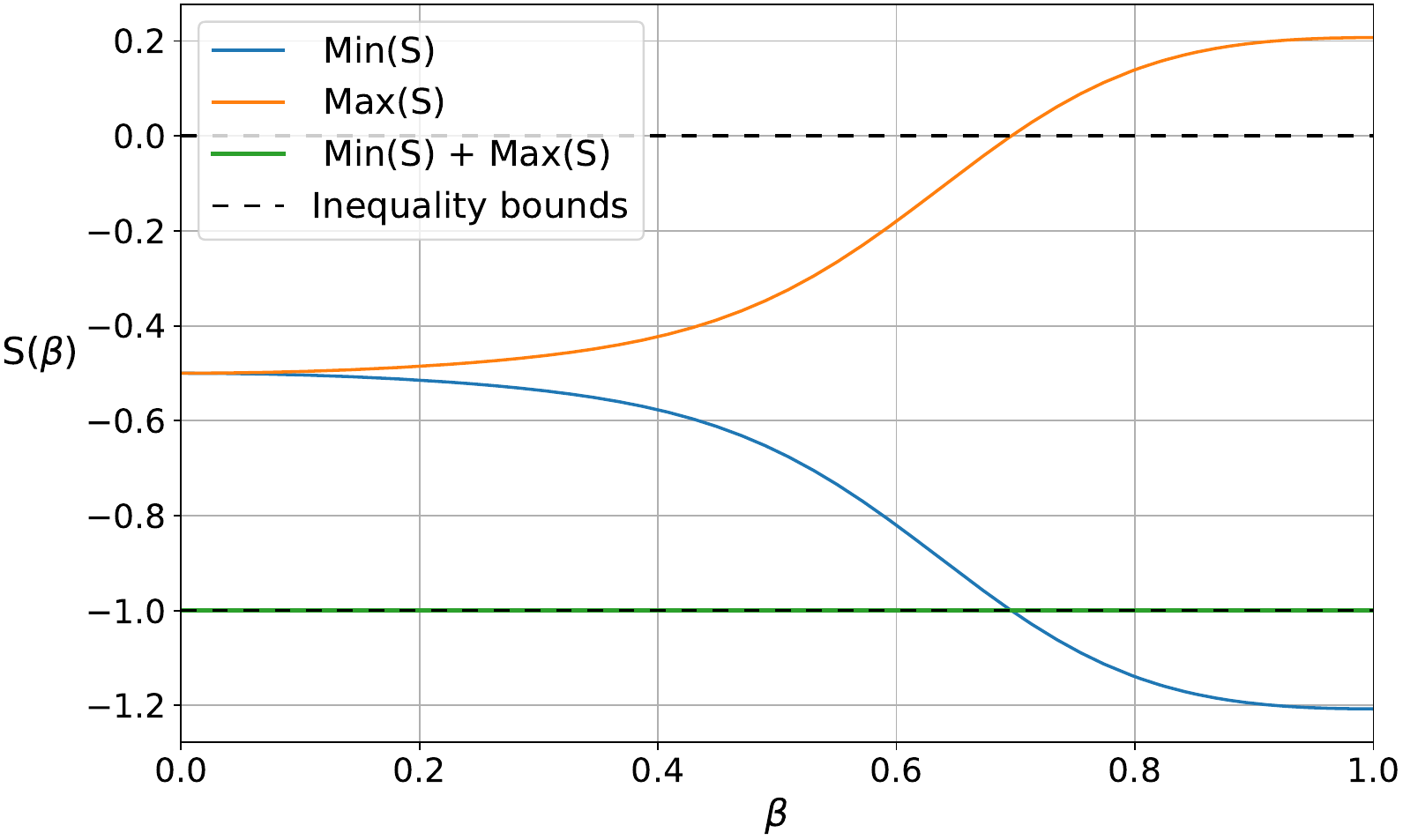}
\centering
\caption{Maximum and minimum values for S($\beta$) in equation (\ref{S_section_2_fully_polarized}). Horizontal dashed lines at $S=0$ and $S=-1$ show the bounds for the CHI.}
\label{S_sec2}
\end{figure}

In order to test the CHI, we calculated the minimum and maximum value of $S$ for each value of $\beta\in [0,1]$, this values were calculated numerically using the \texttt{Wolfram Mathematica} software. The result is shown in Figure \tcolor{\ref{S_sec2}}.\\

Clearly, violation of the CHI occurs for energies such that $\beta \gtrsim 0.696$. The fact that Min$(S)+$Max$(S)=-1$ guarantees that if the CHI can be violated from below for a given $\beta_0$, then it can also be violated from above. Consider $\textnormal{Min}(S) = -1 - \textnormal{Max}(S)$, if a minimum $\textnormal{Min}(S) < -1$ for some $\beta_0$, this implies that $\textnormal{Max}(S) > 0$ for the same $\beta_0$.


\section{INITIALLY UNPOLARIZED PARTICLES}

Again, following the procedure in \tcolor{\cite{yongram}}, for initial positron and electron we consider their momentum in the common CM frame as $\textbf{p}_1 = \gamma m\beta(0,1,0) = - \textbf{p}_2$ respectively and for the final particles $\textbf{k}_1 = \gamma m\beta(1,0,0) = - \textbf{k}_2$. 

\begin{figure}[H]
\includegraphics[scale=0.34]{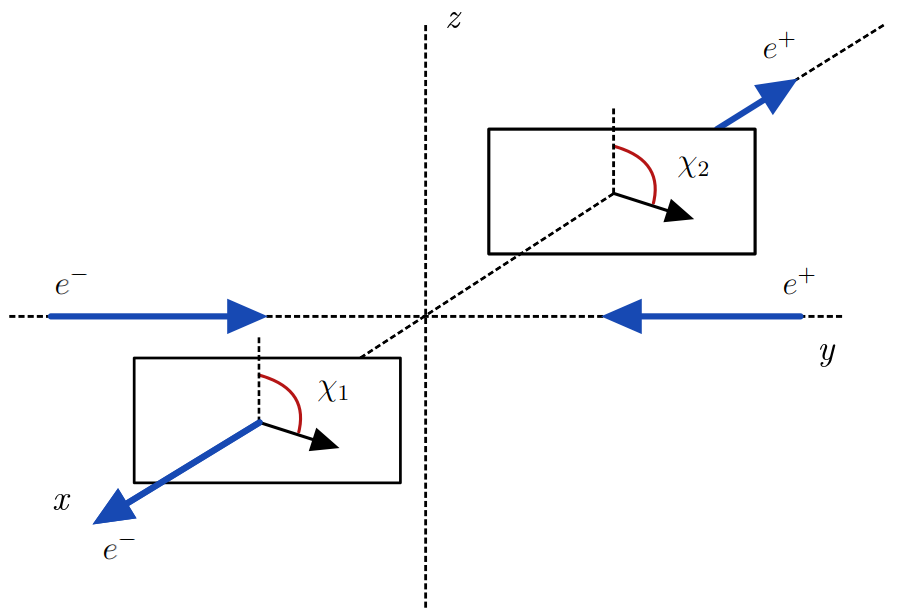}
\centering
\caption{Configuration space scattering diagram. Initial electron and positron move along the $y$ axis, and emergent electron and positron along the $x$ axis. The angle $\chi_1$ is measured relative to the $x$ axis and denotes the electron spin orientation over the $zy$ plane. The angle $\chi_2$ is also measured relative to the $x$ axis and denotes the positron's opposite spin orientation over the $zy$ plane. Remember that for antiparticles spinor direction is opposite to physical particle's spin direction.}
\label{scatter_diagram_initially_unpolarized}
\end{figure}

First we take the amplitude in (\ref{general_amplitude}) and average over initial spins

\begin{align}
|\overline{\mathcal{M}}|^2 = \frac{1}{4}\sum_{s,r}|\mathcal{M}|^2 \spa\spa.
\end{align}

Note that as we previously mentioned, we did not sum over final ones. The squared averaged amplitude yields

{\small	
\begin{align*}
|\overline{\mathcal{M}}|^2 &= 
\frac{e^4}{4s^2}
\Tr\Big[  (\slashed{p_1}+m) \ga_\nu (\slashed{p_2}-m)\ga_\mu  \Big]
\sub{k_1} \ga^\mu \sv{k_2} \svb{k_2 } \ga^\nu \su{k_1}  \\
&+ \frac{e^4}{4t^2} 
\sub{k_1} \ga^\alpha ( \slashed{p_1}+m  ) \ga^\beta \su{k_1}
\svb{k_2} \ga_\beta ( \slashed{p_2}-m  ) \ga_\alpha  \sv{k_2}  \\
&- \frac{e^4}{4st} 
\Tr\Big[(\slashed{p_1}+m)\ga^\sigma\su{k_1}\svb{k_2} \ga_\sigma (\slashed{p_2}-m )\ga_\omega \Big]
\sub{k_1}\ga^\omega \sv{k_2} \\
&- \frac{e^4}{4st} 
\Tr\Big[(\slashed{p_1}+m)\ga_\rho (\slashed{p_2}-m )\ga_\lambda \sv{k_2} \sub{k_1}\ga^\lambda\Big]
\svb{k_2}\ga^\rho\su{k_1} \esp.
\numberthis
\end{align*}
}

This calculations were performed using the spinor expressions shown in appendix B with the help of \texttt{Mathematica}. Because spinors are present inside traces in the interference terms, the gamma matrices algebra cannot help us to perform the calculations. Then we need to adopt a particular representation to perform the calculations explicitly. As authors in \tcolor{\cite{yongram}}, we adopt the Dirac representation. Let $F(\beta,\chi_1,\chi_2)$ be the averaged modulus squared of the amplitude,

\begin{align*}
F(\beta,\chi_1,\chi_2) = \frac{e^4}{4\beta^4} &\phantom{.}
\Bigg[
\cos(\chi_1+\chi_2)  (-\beta^6 + 3\beta^4 - \beta^2)  \\
&+ \cos(\chi_1-\chi_2)  (\beta^6 - 6\beta^4 + 5\beta^2) \\
& - 2\beta^6 +  13\beta^4 - 6\beta^2 + 4
\Bigg]
\spa\spa, \numberthis
\end{align*}

the conditional joint probability of measuring the polarization of the emergent particles in the direction $\chi_1,\chi_2$ is 

\begin{align}
P(\beta,\chi_1,\chi_2) = \frac{F(\beta,\chi_1,\chi_2) }{N(\beta)} \spa\spa,
\end{align}

again, the normalization factor $N(\beta)$ is obtained by summing the non-normalized probability over the pair of angles $\chi_1$ and $\chi_2$, 

\begin{align*}
N(\beta) &= 
F(\beta,\chi_1,\chi_2) 
+ F(\beta,\chi_1 +\pi ,\chi_2) \\
&+ F(\beta,\chi_1,\chi_2 +\pi )
+ F(\beta,\chi_1 +\pi ,\chi_2 +\pi) \\
&= \frac{e^4}{\beta^4} \big(  -  2\beta^6 + 13\beta^4 - 6\beta^2 + 4  \big)
\spa\spa. \label{N(b)_initiall_no_polarized}\numberthis
\end{align*}

The probability of measuring only the polarization of the emerging electron and positron respectively is 

\begin{align}
P(\chi_1 , - ) = \frac{1}{2}  
\spa\spa,\spa\spa
P( - , \chi_2 ) = \frac{1}{2}  \spa\spa.
\end{align}

One way to observe that indeed our calculation is correct is the following: the spin-averaged amplitude for Bhabha scattering at tree level is

\setlength{\jot}{6pt}
\begin{align*} \label{bhabha_scat_fully_avreaged}
\frac{1}{4}\sum_{s,r}\sum_{s',r'}|\mathcal{M}|^2  &= 
\frac{2e^4}{s^2} \big( t^2 + u^2 + 8m^2s - 8m^4 \big) \\
&+ \frac{2e^4}{t^2} \big( s^2 + u^2 + 8m^2t - 8m^4 \big) \\
&+ \frac{4e^4}{st} \big( u^2 - 8m^2u + 12m^4 \big) 
\spa\spa, \numberthis
\end{align*}
\setlength{\jot}{13pt}

where in this case, the Mandelstam variables are

\begin{align}
s = \frac{4m^2}{1 - \beta^2} \enskip\enskip,\enskip\enskip
t = -\frac{2\beta^2 m^2}{1 - \beta^2} \enskip\enskip,\enskip\enskip
u = - \frac{2\beta^2 m^2}{1 - \beta^2} \enskip\enskip.
\end{align}

As is mentioned right after equation (21) in \tcolor{\cite{yongram}}, summing $F(\beta,\chi_1,\chi_2)$ over the pairs of angles $\chi_1$ and $\chi_2$ should be equivalent to sum over polarizations of the emerging particles. If we plug the Mandelstam variables in (\ref{bhabha_scat_fully_avreaged}), we indeed get the normalization factor $N(\beta)$ in equation (\ref{N(b)_initiall_no_polarized})

\begin{align} \label{bhabha_scat_fully_avreaged_evaluated}
&\frac{1}{4}\sum_{s,r}\sum_{s',r'}|\mathcal{M}|^2  = 
\frac{e^4}{\beta^4} \big(  -  2\beta^6 + 13\beta^4 - 6\beta^2 + 4  \big)
\spa\spa,
\end{align}

notice that this can not be obtained in \tcolor{\cite{yongram}}.

\begin{figure}[H]
\includegraphics[scale=0.21]{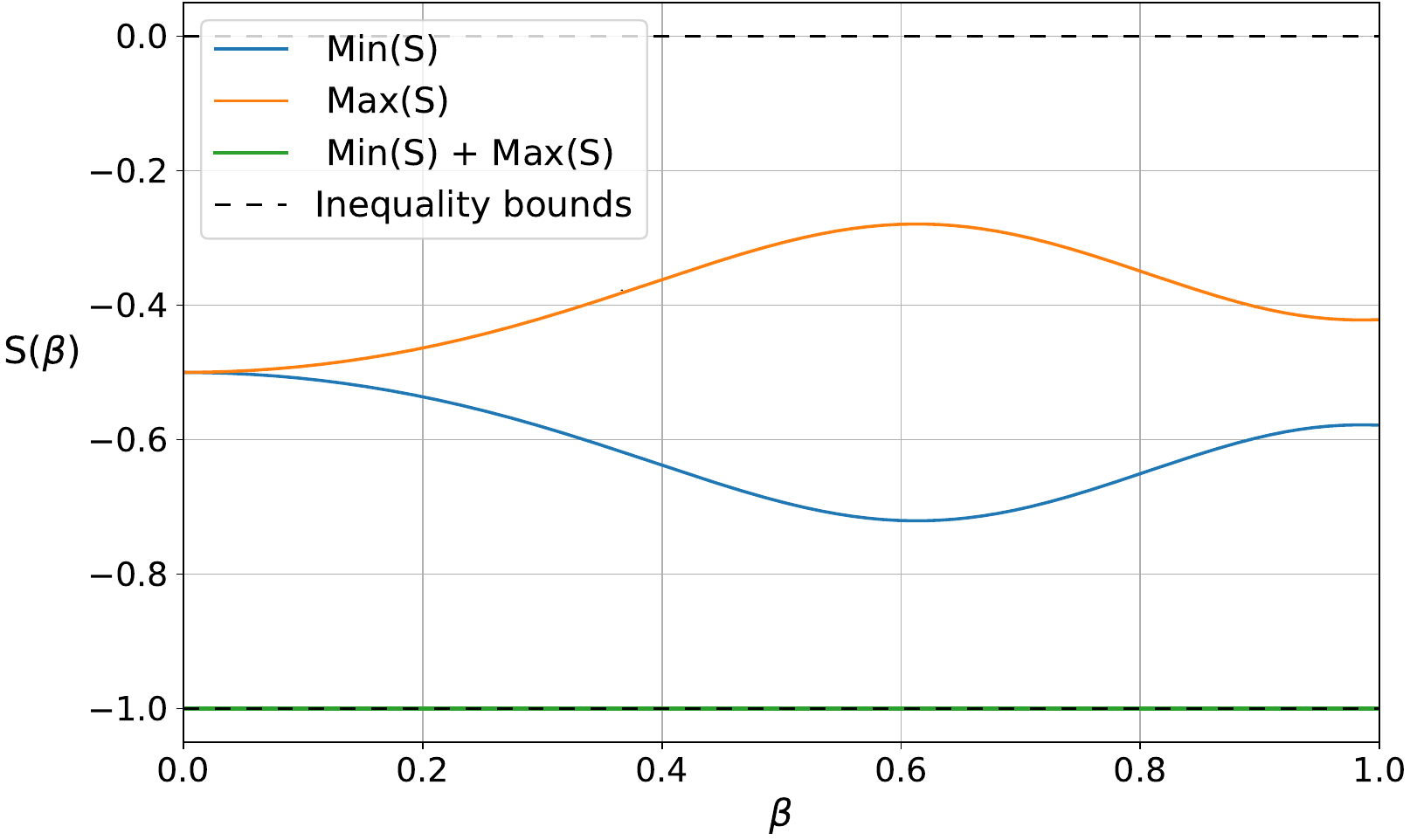}
\centering
\caption{Maximum and minimum values for S($\beta$). Horizontal dashed lines at $S=0$ and $S=-1$ show the bounds for the CHI.}
\label{S_sec3}
\end{figure}

To analyze possible violation of the CHI, again we plot the numerical minimum and maximum value of $S(\beta)$ for each value of $\beta\in [0,1]$. The result is shown in Figure \tcolor{\ref{S_sec3}}. In this case, violation of the CHI do not occur at any energy.

\section{CONCLUSION}

We have tested the Clauser-Horne inequality in the QED process $e^+ e^- \rightarrow e^+ e^-$. First, we considered initially polarized electrons and found violation of the Clauser-Horne inequality for incoming electrons with a speed $\beta \gtrsim 0.696$. As is remarked in \tcolor{\cite{yongram}}, the spin correlation depends on the speed of the particles. As a second case, we studied initial unpolarized electrons, but this time we found no violation of the Clauser-Horne inequality. It is important to remember that entanglement does not necessarily imply violation of Bell-type inequalities. However, if a state violates an inequality, it is assured that the state is entangled. Our result does not imply entanglement is lost due unpolarized particles. As argued in \tcolor{\cite{bell_ineq_at_LEP}}, averaging over initial spin states leads to loss of information on the final states to be measured, thus, is not possible violate the Clauser-Horne inequality this way. Measuring the polarization of the outgoing particles in the scattering process $e^+ e^- \rightarrow e^+ e^-$, where the incoming particles polarization is unknown, is not a proper experimental test of the Clauser-Horne inequality.

\section{ACKNOWLEDGMENTS}

We would like to thank Alfredo Aranda and Carlos Alvarado for useful discussions and guidance on this work.

\section{APPENDIX A: SPINORS FOR SECTION II}

In the Dirac representation of the gamma matrices, the spinors for particles and antiparticles normalized as $\sub{p}^s\su{p}^r = 2m \delta_{sr}$ and $\svb{p}^s\sv{p}^r =- 2m \delta_{sr}$, are given by

\begin{equation} \label{dirac_spinors}
    u^{s}\scaleobj{0.7}{(p)}=\sqrt{E+m} 
    \begin{pmatrix}
    \varphi^s  \\[.27cm]
    \frac{\Vec{\sigma} \cdot \Vec{p}}{E+m} \spa \varphi^s
    \end{pmatrix}
    \spa\spa,
\end{equation}

\begin{equation}\label{dirac_spinors2}
    v^{s}\scaleobj{0.7}{(p)}=-\sqrt{E+m} 
    \begin{pmatrix}
    \frac{\Vec{\sigma} \cdot \Vec{p}}{E+m} \spa \eta^{s} \\[.27cm]
    \eta^{s} 
    \end{pmatrix}
    \spa\spa,
\end{equation}

where $\varphi^s$ and $\eta^s$ are Weyl spinors encoding spin direction. For spin along $x$, $y$ or $z$ axis they are eigenstates of the Pauli sigma matrices $\sigma^j$, if the spin direction is quantized along an arbitrary axis $\hat{n}$, then they are eigenstates of the linear combination $\vec{\sigma} \cdot \hat{n}$.

Working in the center of mass reference frame, lets consider an initial electron with spin up along $z$ axis and four-momentum $p^\mu_1$ and similarly a initial positron also with spin up along $z$ axis and four-momentum $p^\mu_2$, with their momenta given by

\begin{align}
p^\mu_1 = ( E,0,\gamma m\beta,0  )   
\esp\esp&,\esp\esp
p^\mu_2 = ( E,0, -\gamma m\beta,0  )
\spa\spa.
\end{align}

For the electron, physical particle spin is the same as spinor $\su{p_1}^s$ spin. Pointing in the $+z$ direction, the Weyl spinor is

\begin{align} \label{weyl_spinor_phi_1}
\varphi^1 =
\left(\begin{array}{c}
1\\[0.2cm]
0\\
\end{array}\right)
\spa\spa.
\end{align}

For the positron, physical particle spin is opposite to the spinor $\sv{p_2}^r$ spin. For physical spin along the $z$ direction, the corresponding Weyl spinor is

\begin{align} \label{weyl_spinor_eta_2}
\eta^2 =
\left(\begin{array}{c}
0\\[0.2cm]
1\\
\end{array}\right)  \spa\spa,
\end{align}

plugging this into (\ref{dirac_spinors}) we get the initial spinors 

\begin{align}
\su{p_1} = \sqrt{E + m}
\left(\begin{array}{c}
1  \\[0.1cm]
0  \\[0.1cm]
0  \\[0.1cm]
i\rho
\end{array}\right)
\spa\spa,\spa\spa
\sv{p_2} = - \sqrt{E + m}
\left(\begin{array}{c}
i\rho  \\[0.1cm]
0  \\[0.1cm]
0  \\[0.1cm]
1
\end{array}\right)
\spa.
\end{align}

Now for the final electron and positron with momenta $k^\mu_1$ and $k^\mu_2$ respectively 

\begin{align}
k^\mu_1 = ( E,0,0,\gamma m\beta  )   
\esp\esp&,\esp\esp
k^\mu_2 = ( E,0,0,-\gamma m\beta  )
\spa\spa,
\end{align}

we look for spinor spin lying on the $xy$ plane measured relative to the $x$-axis for both particles. For this we take the Weyl spinor in (\ref{weyl_spinor_phi_1}) and rotate it using the spin rotation operator 

\renewcommand\arraystretch{2.2}
\begin{align}\label{Rotation}
\eu^{-i\frac{\theta}{2} \vec{\sigma} \cdot \hat{n}} =
\scaleobj{0.8}{
\left(\begin{array}{cc}
\cos(\theta/2)-in^{3}\sin(\theta/2)    \esp&\esp   -i(n^{1}-in^{2})\sin(\theta/2)  \\
-i(n^{1}+in^{2})\sin(\theta/2)         \esp&\esp   \cos(\theta/2)+in^{3}\sin(\theta/2)
\end{array}\right)
}
\esp.
\end{align}
\renewcommand\arraystretch{1}

Rotations are performed by an angle $\theta$ around the $\hat{n}$ axis. As illustrated in Figure \tcolor{\ref{scatter_diagram_initially_unpolarized}}, looking at the electron like coming towards us, spin direction is measured clockwise, looking at the positron like coming towards us, spin direction is also measured clockwise so the Weyl spinor is the same for both spinors $\su{k_1}$ and $\sv{k_2}$. The physical positron spin direction measured in the lab is opposite to the spinor $\sv{k_2}$ direction, so for a $\chi_2$ spinor spin direction, the lab measures a $\chi_2 + \pi$ spin direction. First rotate $\varphi^1$ by an angle $\theta = \pi/2$ around the $y$-axis, then rotation again this time y an angle $\chi_1$ around the $z$ axis. Then repeat for the positron but perform the second rotation by an angle $\chi_2$, the Weyl spinor is

\begin{align}
\xi_k = \frac{1}{\sqrt{2}}
\left(\begin{array}{c}
\eu^{-i\chi_k /2}  \\[0.25cm]
\eu^{-i\chi_k /2}
\end{array}\right)
\spa\spa.
\end{align}

Plugging this into (\ref{dirac_spinors}) we obtain the spinor expressions for the final particles

\begin{align}
\su{k_1} = \sqrt{\frac{E+m}{2}}
\left(\begin{array}{c}
\eu^{-i\chi_1 /2}  \\[0.15cm]
\eu^{i\chi_1 /2}  \\[0.15cm]
\rho \spa \eu^{-i\chi_1 /2}  \\[0.15cm]
-\rho \spa \eu^{i\chi_1 /2}  \\[0.15cm]
\end{array}\right)
\spa\spa,
\end{align}

\begin{align}
\sv{k_2} = - \sqrt{\frac{E+m}{2}}
\left(\begin{array}{c}
- \rho \spa \eu^{-i\chi_2 /2}  \\[0.15cm]
\rho \spa \eu^{i\chi_2 /2}  \\[0.15cm]
\eu^{-i\chi_2 /2}  \\[0.15cm]
\eu^{i\chi_2 /2}  \\[0.15cm]
\end{array}\right)
\spa\spa.
\end{align}

\section{APPENDIX B: SPINORS FOR SECTION III}
Consider the final electron and positron as having momenta $k^\mu_1 = ( E,\gamma m\beta,0,0)$ and $k^\mu_2 = ( E,-\gamma m\beta,0,0)$, and suppose their spins lie on the $yz$-plane. To describe the Weyl spinors in (\ref{dirac_spinors}) and (\ref{dirac_spinors2}), it is enough to rotate $\varphi^1$ from (\ref{weyl_spinor_phi_1}) around the $y$-axis. Denoting as $\chi_1$ and $\chi_2$ the rotation angles that describe the spin direction of the electron and positron, and using the expression (\ref{Rotation}), we obtain the Weyl spinor

\begin{align}
\xi_k =
\left(\begin{array}{c}
\cos{(\chi_k /2)}  \\[0.2cm]
-i\sin{(\chi_k /2)}
\end{array}\right)
\spa\spa.
\end{align}

Plugging this expression into (\ref{dirac_spinors}) and (\ref{dirac_spinors2}), we obtain the final four-spinors:

\begin{align}
\su{k_1} = \sqrt{E+m}
\left(\begin{array}{c}
\cos{(\chi_1 /2)}  \\[0.13cm]
-i\sin{(\chi_1 /2)}  \\[0.13cm]
-i \rho \spa \sin{(\chi_1 /2)}  \\[0.13cm]
\rho \spa \cos{(\chi_1 /2)}  \\[0.13cm]
\end{array}\right)
\spa\spa,
\end{align}
\begin{align}
\sv{k_2} = - \sqrt{E+m}
\left(\begin{array}{c}
i \rho \spa \sin{(\chi_2 /2)}  \\[0.13cm]
-\rho \spa \cos{(\chi_2 /2)}  \\[0.13cm]
\cos{(\chi_2 /2)}  \\[0.13cm]
-i \sin{(\chi_2 /2)}  \\[0.13cm]
\end{array}\right)
\spa\spa.
\end{align}

\renewcommand*{\bibfont}{\footnotesize}
\renewcommand{\refname}{\centering\textsc{REFERENCES}}
\bibliography{bib}

\end{multicols}

\end{document}